\documentstyle[12pt]{article}

\newcommand{\eq}[1]{Eq.(\ref{#1})}
\def\ctg{\mbox{ctg}}
\def\arctg{\mbox{arctg}}
\def\be{\begin{equation}}
\def\ee{\end{equation}}
\def\a{\alpha}
\def\b{\beta}

\def\s{\sigma}
\def\r{\rightarrow}
\def\UU{\tilde{U}}
\def\e{\epsilon}
\def\G{\Gamma}
\def\cc{\tilde{c}}
\def\ck{c^{+}}
\def\dk{d^{+}}
\def\bk{b^{+}}

\begin{document}
\vspace{1.5in}
\begin{center}

{\Large \bf Metal - Insulator transition in the  \\
Generalized Hubbard model  } \\
\vspace{0.4in}
{\large \bf Anatoly~ A.~Ovchinnikov}   \\
\vspace{0.2in}
{\it Institute for Nuclear Research of the Russian Academy
of Sciences, \\ Moscow 117312 Russia} \\
\vspace{0.85in}
{\bf Abstract} \\
\end{center}

We present the exact ground-state wave function and energy of
the generalized Hubbard model, subjected to the condition that
the number of double occupied sites is conserved,
for a wide, physically relevant range of parameters.
For one hole and one double occupied site the existence of the
ferromagnetic ground-state is proved which allow one to determine the
critical value of the on-site repulsion corresponding to the
point of metal-insulator transition.
For the one dimensional model the exact solution for special
values of the parameters is obtained.

\newpage

               {\bf 1.~Introduction}
\vspace{0.15in}

The Hubbard model is the generic model to describe correlations in
narrow-band systems \cite{H}. The on-site repulsion is due to the
matrix elements of the Coulomb interaction corresponding to the on-site
Wannier states while the other matrix elements are neglected. However the
on-site repulsion can be sufficiently strong and the values of the
matrix elements corresponding to the pair of nearest-neighbor sites
can be comparable with the value of the simplest nearest-neighbor
hopping amplitude. The Hamiltonian of the model which is often referred to
as a generalized Hubbard model contains the interaction terms
of the fourth order in the electron creation and annihilation operators
corresponding to the nearest-neighbor sites. The Hamiltonian is
\[
\hat{H}=-t\sum_{<ij>\s}( c_{i\s}^{+}c_{j\s} + c_{j\s}^{+}c_{i\s})+
X\sum_{<ij>\s}(c_{i\s}^{+}c_{j\s}+c_{j\s}^{+}c_{i\s})(n_{i-\s}+n_{j-\s})
\]
\be
+V\sum_{<ij>} n_i n_j + U\sum_{i} n_{1i} n_{2i},
\label{o}
\ee
where $\s=1,2$ - is the projection of spin,
$n_{i\s}=c_{i\s}^{+}c_{i\s}$, $n_i=n_{1i}+n_{2i}$
and $<ij>$ denotes the pair of nearest neighbor sites.
The generalized model (\ref{o}) has been studied previously
by several authors \cite{HM,VS,Others}.
The other models with a similar kind of
hopping term where considered in ref's \cite{Korepin,MR}.
Even for $X,V \ll U$ the presence of interactions which directly couple
nearest-neighbor sites should lead to new effects.
For example the correlated hopping terms are believed to play an essential
role in the formation of high -$T_c$ superconductivity \cite{HM}.
In the present letter we consider the Hamiltonian (\ref{o}) at $X=t$.
At this particular value the model is much more simple than the
conventional Hubbard model.
Note that the corresponding region in space of parameters is quite
realistic in view of the estimates of these parameters for different systems
(for example, see \cite{VS,Others} and references therein).
At the half filling the ground state of the model
at $X=t$ can be found exactly in any dimensions in a wide range of the
parameters $t,U,V$. We also study the metal-insulator transition at the
critical value $U_c$ which can be found exactly in our case
in contrast to the Hubbard model where the Mott picture of
metal-insulator transition \cite{Mott} is not directly applicable
at least for a simple square or cubic lattice where presumably
(in spite of the predictions based on the Gutzwiller approximation
\cite{BR}) the system is the antiferromagnetic insulator at arbitrary $U$.
Recently the same model was studied by Strack and Vollhardt \cite{VS}
with the help of supersymmetric representation.
We show that the wave functions proposed are in fact the ground states
of the model in a range of parameters which is much wider than the
region found in ref.\cite{VS}.

First, we show that at the half filling the exact ground state of the
model at $X=t$ and $U>z\max(2t,V)$ ($z$ is the coordination number of
the lattice) is a highly degenerate state without
the double occupied sites and the system is paramagnetic insulator.
For the on-site repulsion $U<U_c$, where $U_c$ is the critical value,
which is not necessarily coincide with the obtained bound, the creation
of holes and double occupied sites is energetically favorable.
At this point the transition to the metallic state take place.
In complete analogy with the Nagaoka theorem for the infinite-U Hubbard
model \cite{N} we prove that in the sector of Hilbert space with one hole
and one double occupied site the state with the lowest energy is
ferromagnetic.
We show that at $V<2t$ the critical value is determined by the ground
-state energy of the problem with one hole and one double occupied site
and $U_c=2zt$.
At present time the stability of the
Nagaoka state at finite concentration of holes is not proved. However
it is supposed to be the correct ground state at
sufficiently small concentration of holes \cite{A}.
Assuming the stability of the ferromagnetic state at finite
concentration of holes and double occupied sites at low concentration
of holes we find the
density of holes below the point of metal-insulator transition.
For a bipartite lattice we find another region of the parameters
where the determination of the ground state is possible.
For a square or cubic lattice at $U<2zV-z\max(2t,V)$ the ground state is
given by the state with the electrons occupying only one of the sublattices.
In particular, at $V>2t$ the ground state is known exactly at arbitrary
$U$. The transition between two different ground states take place at
$U=zV$.
Finally we consider the generalized Hubbard model at $X=t$ in
one dimension. The dependence  of the energy on the total spin for an
arbitrary number of holes is studied.
These results may be useful in the context of study of the
stability of the ferromagnetic state for the infinite U Hubbard model
in higher dimensions.
It is shown that the model is exactly solvable at $V=0$.
We also comment on the behavior of
different generalizations of the model (\ref{1}) which include
the antiferromagnetic coupling.
Brief description of the results presented in section 2 was given
in ref.\cite{Ovch}.

\vspace{0.25in}
         {\bf 2.~Generalized Habbard model in arbitrary dimensions}
\vspace{0.15in}

Consider the model (\ref{o}) at the half-filling
($\bar{n}=\bar{n}_1+\bar{n}_2=1$) at $X=t$. The Hamiltonian
\be
\hat{H}=-t\sum_{<ij>\s} c_{i\s}^{+}c_{j\s} (1-n_{i-\s}-n_{j-\s})+h.c.
\label{1}
\ee
\[
+ V \sum_{<ij>} n_i n_j + U \sum_{i} n_{1i} n_{2i},
\]
where $h.c.$ stands for hermitean conjugate, conserves the
number of double occupied sites
$\hat{N}=\sum_{i}n_{1i}n_{2i}$: $[\hat{H},\hat{N}]=0$.
The eigenstates of $\hat{H}$
corresponds to a definite number of double occupied sites $N$ which at the
half filling coincide with the number of holes. First, let us prove that
at $U>z\max(2t,V)$ the ground state corresponds to $N=0$. It is convenient
to express the Hamiltonian (\ref{1}) in terms of the fermionic
creation and annihilation operators of the holes ($c_i^{+},c_i$) and the
double occupied sites ($d_i^{+},d_i$) defined starting from the
ferromagnetic state $|F>=\sum_{i}c_{2i}^{+}|0>$.
The up-spin electrons are described by the
Holstein-Primakoff hard-core bose operators ($b_i^{+},b_i$).
To obtain the interaction term $\sim V$ one can make the following
substitution:
\[
n_{1i}=n_i + n_{di}, ~~~~~~~n_{2i}=1-n_i-n_{ci}.
\]
The Hamiltonian is
\be
\hat{H} = -t\sum_{<ij>}(c_{i}^{+}c_{j} + d_{i}^{+}d_{j})
  (b_j^{+}b_i+(1-n_i)(1-n_j)) + h.c.
\label{2}
\ee
\[
 + V \sum_{<ij>} (n_{ci}-n_{di})(n_{cj}-n_{dj}) + U\hat{N} + zVL/2,
\]
where
$n_{ci}=c_i^{+}c_i,~ n_{di}=d_i^{+}d_i,~n_i=b_i^{+}b_i$,
and the constraint $n_{ci}+n_{di}+n_i\leq 1$ which is equivalent to
the infinite on-site repulsion between the particles is implied.
$\hat{N}=\sum_{i}n_{di}$ is the number of double
occupied sites and the energy $NU$ is due to the last term in \eq{1}.

{\bf a) Variational theorem.}  The
upper bound for the ground-state energy $E_0 \leq zVL/2$ ($L$ is the number
of lattice sites) of the Hamiltonian (\ref{1}) can be obtained using the
variational wave function with singly occupied sites
\be
|\phi>=\prod_{i\in {\cal L}}c_{1i}^{+}
\prod_{j\in {\cal L}^{\prime}}c_{2j}^{+}|0>,
\label{var}
\ee
where ${\cal L}$ and ${\cal L}^{\prime}$ are arbitrary disjoint
sets of lattice sites which together build up the total lattice.
In the representation (\ref{2}) $|\phi>$ is the state without the fermions.
It is easy to derive the lower bound for the energy $E_0$.
The Hamiltonian (\ref{2}) can be considered as a matrix $H_{\a\b}$
where the indices $\a,\b$ are enumerate the set of possible configurations
of particles on the lattice
$\a=(i_1...i_N|j_1...j_N|l_1...l_M)$ where $i,j,l$ are the coordinates
of holes, double occupied sites and hard-core bosons respectively
(see Appendix).
One can see that due to the Fermi statistics of $c$- and $d$- particles the
non-diagonal matrix elements of $H_{\a\b}$ corresponding to the kinetic
energy term of the Hamiltonian (\ref{2}) are equal to $\pm t$.
Clearly, for the Bose statistics these matrix elements would be equal to $-t$.
The diagonal matrix elements are determined by the second term of \eq{2}.
The following theorem can be easily proved. For each of the eigenvalues
$E$ of any hermitean matrix $H_{\a\b}$  at least one of the inequalities
\be
|E-H_{\a\a}| \leq \sum_{\b\neq\a}|H_{\a\b}|
\label{Haa}
\ee
is satisfied. In particular, for the restriction on $E_0$ one should take
the minimal value of $H_{\a\a}$ ($-zVN$). The right-hand side of this
inequality is determined by the number of hopping processes allowed
for a given state $\a$. Since the total number of $c$- and $d$ - particles
is $2N$ the maximal value of the right-hand side of \eq{Haa} is $2ztN$.
In this way we find the following lower bound for the ground-state
energy: $E_0(N)>(-2zt-zV+U)N+zVL/2$.
One can further improve this estimate in the following way.
For example, the hopping of $c$- particle to the nearest
neighbor site occupied by $d$- particle is not possible and one should not
take into account these terms in the right-hand side of \eq{Haa}.
However in this case the contribution
$-V$ to the diagonal matrix element
$H_{\a\a}$ does exist.
In the opposite case of isolated $c$ and $d$- particles the binding energy
is absent while the hopping processes are possible. Thus we obtain the
following lower bound for the energy:
\be
E_0(N) \geq (- z \max(2t,V) + U) N + zVL/2.
\label{lower}
\ee
Taking into account the upper bound $E_0<zVL/2$  and \eq{lower}
we see that at
$U\geq z\max(2t,V)$ the ground-state wave function does not contain the
holes and the double occupied sites, $N=0$, and the energy is exactly equal
to $E_0=zVL/2$. The ground state is $2^L$-fold degenerate and is given
by \eq{var}. For comparison the bound for U found in ref.\cite{VS} is
$4zt+zV$.
The authors use the following method to obtain the lower bound
for the energy. With the help of the operators
\[
 P_{ij\s}^{+}=c_{i\s}(1-n_{i-\s})+c_{j\s}(1-n_{j-\s}),~~~\,
 Q_{ij\s}^{+}=c_{i\s}^{+}n_{i-\s}+c_{j\s}^{+}n_{j-\s},
\]
the Hamiltonian can be represented in the form
\[
   \hat{H}=t\sum_{<ij>\s}(P_{ij\s}^{+}P_{ij\s}+Q_{ij\s}^{+}Q_{ij\s})
   -2zt\sum_{i}(1-n_i)
\]
\be
   +(U-4zt)\sum_{i}n_{1i}n_{2i}+V\sum_{<ij>}n_{i}n_{j}.
\label{PQ}
\ee
Since the average of the first term in \eq{PQ} over the ground state
is positive-definite the bound $U>4zt+zV$ is obtained.
Clearly for an arbitrary lattice, at
$t,V\ll U$ and a small deviation $t-X\neq 0$ the degeneracy is absent
and the model (\ref{o}) reduces to an effective Heisenberg model with
antiferromagnetic coupling constant $4(t-X)^2/U$. For a bipartite
lattice one has the antiferromagnetically ordered ground state.

For $U<U_c$ the creation of holes and double occupied sites take place.
We shall see that at $V<2t$
the critical value $U_c$ is determined by the value of the
lowest energy for the state with
one empty and one double occupied site ($N=1$).
This problem can be solved exactly in two and three dimensions.
The Hamiltonian (\ref{2}) has the same form as an analogous
Hamiltonian for the infinite-$U$ Hubbard model away from the half filling.
The only difference is the absence of $d$- fermions in the latter case.
The hopping term for the bosons is absent as well and their interaction
with the fermions has the same form.
According to the Nagaoka theorem \cite{N}, for one hole the ground
state of the infinite-$U$ Hubbard model is ferromagnetic (the total spin
is maximal). In our case it is possible to prove that at $N=1$ i.e.
for one hole and one double occupied site the ground-state is
ferromagnetic.
Following the original Nagaoka proof consider the hopping process
in which the hole and the double occupied site starts from a given
positions and come back to the same positions after a number of steps.
The configuration of spins (the bosons) can be different in the initial and
the final states. For one hole the corresponding energy-dependent
self-energy part introduced in ref.\cite{N} is positive.
The same would be true for arbitrary number of holes if the holes
would obey the Bose statistics in a sense of the Hamiltonian (\ref{2}).
Clearly in our case the positivity condition is satisfied since for $N=1$
the statistics of $c$- and $d$- particles which represent the two
different species of fermions is not important.
The other steps of the Nagaoka proof can be applied without
modification (the presence of the attraction $\sim V$ in \eq{2} is
either not important).
Thus at $N=1$ the state with $S=S_{max}-1$ which is the maximal total
spin for a state with one hole is the ground state of the system.
In the representation (\ref{2}) that means that the number
of bosons is equal to zero for the projection of spin $S^z=-(S_{max}-1)$.
The ground-state energy of two particles has the form
$E_0(N=1)=-2zt+U+a/L$, where the last term is due to the interaction of
$c$- and $d$- particles.
Let us suppose that the interaction part of the energy is positive $a>0$
(the corresponding condition for the parameter $V$ will be found later).
Then the critical value of $U$ is
\be
  U_c = 2zt.
\label{crit}
\ee
At $U>U_c$ the ground state is given by \eq{var} and at $U<U_c$
the creation of holes is energetically favorable. The critical value
(\ref{crit}) is the point of metal-insulator transition.

Before solving the two-particle problem let us comment on the
behavior of the system at finite density of $c$- and $d$- fermions.
At present time the stability of the
Nagaoka state for the infinite- repulsion Hubbard model
at finite concentration of holes is not proved.
It is supposed that the ground state is ferromagnetic at
sufficiently small concentration of holes (for discussion see ref.\cite{A}).
In our case we will also assume the stability of the ferromagnetic state
($S=S_{max}-N$)
at sufficiently small density of holes and double occupied sites
$\rho=N/L$.
Then at $U<2zt$ and $|2zt-U|\r 0$ the density $\rho\r 0$ and the analog of
the Nagaoka state is realized.
The system of two species of interacting fermions at equal
density $\rho$ should be considered.
At low density the ground-state energy can be evaluated as a series in
the small parameter ($|\ln\rho|^{-1}$ and $\rho^{1/3}$ respectively in
two and three dimensions).
At low density the dependence of the energy on $\rho$ has the form
\be
   E/L= - (2zt-U)\rho + {\cal E}_0 (\rho)
           + \frac{1}{2}a(\rho)\rho^{2},
\label{E}
\ee
where the second term is due to the Fermi statistics. In the lowest order
${\cal E}_0(\rho)=4\pi t\rho^2$ in two dimensions and
${\cal E}_0(\rho)=({1\over 5}6^{5/3}\pi^{4/3})t\rho^{5/3}$
in three dimensions.
In the lowest order in $\rho$ the function $a(\rho)$ is determined
by the two-particle scattering amplitude at low energy.
The scattering amplitude can be expressed through the interaction energy
of two particles of different species $a/L$ in the finite volume $L$.
In fact the low density limit of $a(\rho)$ coincide with the value
of the parameter $a$.
The behavior of the function $a(\rho)$ is different in two and three
dimensions. In three dimensions it is the constant
in the lowest order in $\rho$ while in two dimensions
$a(\rho)=8\pi t/|\ln\rho|+O(t/(\ln\rho)^2)$ \cite{Bloom}.
The ground-state density of holes is determined from the condition
of minimum of the expression (\ref{E}). In the limit $U\r 2zt$
the minimum is determined by the first two terms in \eq{E}.
For example, in two dimensions
\be
\rho_0 = \frac{2zt-U}{4\pi}
\label{rho0}
\ee
with the accuracy up to the terms of order $\sim 1/\ln\rho_0$.
One can also take into account the leading corrections in $\rho_0$
in this formula.
$\rho_0$ is small at small deviation of U from the critical value
(\ref{crit}) in agreement with our assumption about the stability of
the ferromagnetic state. Although the description of the system (\ref{2})
at arbitrary number of bosons is not possible it will be shown that
at $2zt-U\r 0$ the minimum of the energy as a function of $\rho$ found from
\eq{E} is the correct ground-state energy of the system in agreement with
\eq{crit}.

{\bf b) Solution of the two-particle problem.}
Let us proceed with the solution of the two-particle problem.
The interaction potential (\ref{2}) contains the infinite on-site
repulsion $\UU\r\infty$ and the attraction of strength V at
nearest neighbor sites. Let us consider the case of three dimensional
cubic lattice. The ground state corresponds to the total momentum of two
particles equal to zero. One can seek for the wave function in the form
\be
  |\psi>=\sum_{k} F(k) c_k^{+}d_{-k}^{+}|0>,
\label{two}
\ee
where $c_k^{+}=L^{-1/2}\sum_{i}e^{ik{\bar i}}c_i^{+}$.
The function (\ref{two}) is the eigenvector of eigenvalue $E$
if the function $F(k)$ satisfies the Shrodinger equation
\be
(E-2t\e_p)F(p)=\frac{1}{L}\sum_{k}\left(\UU - V\e_{k-p}\right) F(k),
\label{F}
\ee
where $\e_k=-2\sum_{\a=1}^{3}\cos k_{\a}$.
Let us define the function $J(k)=(E-2t\e_k)F(k)$. The equation
(\ref{F}) takes the form
\be
J(p)=\frac{1}{L}\sum_{k} \frac{\UU -V\e_{k-p}}{E-2t\e_k} J(k).
\label{J}
\ee
In order to solve the equation (\ref{J}) for the ground-state energy
$E_0=-2zt+a/L$ let us extract from the sum
the term with $k=0$, which is most divergent for $L\r\infty$. We can
also substitute the value $2t\e_0$ ($\e_0=-z$) for E in the sum over
$k\neq 0$ in \eq{J} since in three dimensions the energy difference
$\e_k-\e_0$ is at least of order $\sim L^{-2/3}$ and the
interaction correction is of order $\sim 1/L$.
We get
\be
J(p) = \frac{(\UU-V\e_p)J(0)}{2ta} +
\frac{1}{L}\sum_{k\neq 0} \frac{\UU -V\e_{k-p}}{2t(\e_0-\e_k)} J(k).
\label{x}
\ee
Defining the new function $\G(k)=aJ(k)/J(0)$ we obtain from (\ref{x})
the equation
\be
2t \G(p) = \UU-V\e_p +
\frac{1}{L}\sum_{k\neq 0} \frac{\UU -V\e_{k-p}}{\e_0-\e_k} \G(k),
\label{G}
\ee
which is nothing else but the equation for the scattering amplitude at
zero energy $\G(0)=a$. The sum can be replaced by the integral in \eq{G}.
The solution of the equation (\ref{G}) can be represented in the form
$\G(k)=\G_0+\e_k\G_1$. Substituting this function into \eq{G} we get
two equations for the unknown constants $\G_0,\G_1$. In the limit
$\UU\r\infty$ the result of the calculations for the simple cubic lattice
($z=6$) has the form
\be
  a = \frac{z(2t-V)}{ Wz-V(Wz-1)/2t },
\label{a}
\ee
where $W=0.2527$ stands for the Watson integral
\[
 W = \frac{1}{2(2\pi)^3}
\int^{\pi}_{-\pi}\int^{\pi}_{-\pi}\int^{\pi}_{-\pi} dk_x dk_y dk_z
\frac{1}{3-\cos k_x -\cos k_y -\cos k_z}.
\]
The expression (\ref{a}) is valid at $V<2t$. At $V=2t$
the amplitude $a$ vanishes which indicate the existence of the
two-particle bound state at $V>2t$.
Vanishing of the interaction correction to the energy at $V=2t$ can be
seen from analogy with the two-magnon problem in the ferromagnet
where in the ground state the total spin should be maximal.
The same conclusion can be made for two dimensions.
To calculate the parameter $a$ one should substitute the sum
\[
\frac{1}{L}\sum_{k\neq 0} \frac{1}{\e_k-\e_0}
\sim \frac{1}{4\pi} \ln L
\]
for W in the formula (\ref{a}). The corrections to the equation
(\ref{G}) are of order $(1/\ln L)^2$ and the terms of that order in
\eq{a} cannot be fixed. However the expansion in $1/\ln L$ breaks
down only at $2t-V\sim 1/\ln L$ and at these values of V the perturbation
theory in $2t-V$ can be used since at $V=2t$ the ground-state wave
function is known:
$\hat{P}_G c_0^{+} d_0^{+}|0>$ (the analog of $S=S_{max}$ state in
the ferromagnet; $\hat{P}_G$ is the Gutzwiller projector).
As in 3D the interaction correction changes sign at $V=2t$ and the
bound-state solution appears at $V>2t$.

{\bf c) Formal proof of the equation (\ref{crit}).}
Finally it is necessary to show that, whether or not the Nagaoka state
is realized at a given density $\rho$, the correction to the energy,
\be
E/L= -(2zt-U)\rho + {\cal E}(\rho),
\label{formal}
\ee
is strictly positive ${\cal E}(\rho)>0$ and does not vanish in the
thermodynamic limit at $\rho\neq 0$.
That means that the density is really
small in the neighborhood of the point of
metal-insulator transition. We have to obtain the lower bound for
${\cal E}(\rho)$ (\ref{formal}).
Let us modify the Hamiltonian (\ref{2}) in such a way
in order to decrease the corresponding ground-state energy.
First, let us replace in \eq{2} the repulsion
$V\sum(n_{ci}n_{cj}+n_{di}n_{dj})$ at nearest neighbor sites by the
attraction of the same form ($V\to-V$) and then make the substitution
$V\to 2t$ so that the resulting interaction takes the form
\[
-2t\sum_{<ij>}(n_{ci}+n_{di})(n_{cj}+n_{dj}) + U\hat{N}.
\]
Second, instead of fermions, consider the particles $c,d$,
obeying the Bose statistics. Since for bosons the Nagaoka state is
the ground state at arbitrary density and the Hamiltonian is symmetric
in respect to the replacement $c\leftrightarrow d$,
the ground-state wave function
$\phi(i_1...i_N|j_1...j_N)$ which is the totally symmetric function
of its arguments coincide with the ground- state wave function
of the Heisenberg ferromagnet in the representation of the
Holstein-Primakoff bosons with the projection of spin $S^z=S_{max}-N$
and $S=S_{max}$.
In fact, the corresponding wave function, which is the
positive-definite and the totally symmetric function, is the
eigenstate of the modified Hamiltonian. Actually it is the ground-state of
the modified Hamiltonian since for the wave function which changes
sign the substitution
\[
\phi(i_1...i_N|j_1...j_N)\rightarrow |\phi(i_1...i_N|j_1...j_N)|
\]
would lower the energy and the positive-definite eigenstate is unique
because of the orthogonality condition.
Note that these considerations can be a basis of a simple proof of the
Nagaoka theorem both in our case and in the case of the infinite-U
Hubbard model \cite{N}.
Therefore the lower bound for the energy is $(-2zt+U)N$
and we find that ${\cal E}(\rho)>0$. Clearly at finite density
${\cal E}(\rho)$ does not vanish in the thermodynamic limit.
Thus it is proved that at $2zt-U\to 0$
the value of $\rho$ minimizing the energy $\rho_0\to 0$.
Consequently, at $V<2t$
the point of metal-insulator transition is indeed given by \eq{crit}.

According to the Mott picture for large coupling $U$ the density of states
exhibits two bands with the centers separated by $U$. In the absence of
electron correlations the width of each Hubbard band is $zt$, and the
gap between the bands is expected to vanish at $U/2zt=1$.
The Hubbard bands are usually obtained in the framework of a special
single - particle Green function decoupling approximation scheme
proposed by Hubbard \cite{H}, the Hartree-Fock - type approximation
which was exact for zero interaction energy or zero band-width
(which amounts to a specific decomposition of a certain Green functions
and has no justification; for example see ref.\cite{Cyrot} for a critical
discussion).
However, strictly speaking, in the
conventional Hubbard model the very notion of the Hubbard bands is
justified only in the limit $U\to\infty$, since the number of double
occupied sites is conserved only in this limit.
In our model the number
of double occupied sites is conserved and the notion of the Hubbard
bands has a precise meaning beyond the framework of any approximation.
Note also that our results can be used to explain
the results of the numerical calculations for a small lattices \cite{Raedt}.

{\bf d) Ground state for a bipartite lattice.}
For an arbitrary $bipartite$ lattice one can find another region of the
parameters where the determination of exact ground state is possible.
For example consider the simple square or cubic lattice. The wave function
which corresponds to a charge-density wave with maximal order parameter,
\be
     |\chi> = \prod_{i\in {\cal A}} c_{1i}^{+} c_{2i}^{+}|0>,
\label{v}
\ee
where ${\cal A}$ is one of the sublattices, is an eigenfunction
of $\hat{H}$. The wave function (\ref{v}) can be used to obtain
an upper bound for the energy: $E_0\leq UL/2$.
To obtain the lower bound it is convenient to define the operators
$\cc_{1i}=(-1)^i c_{1i}$, $\cc_{2i}=(-1)^i c_{2i}^{+}$. At the half
filling the particle number corresponding to the new operators is the same.
In terms of the operators $\cc_{i\s},\cc_{i\s}^{+}$ the state $|\chi>$ is
an antiferromagnetically ordered state with singly occupied sites.
Since in terms of $\cc_{i\s},\cc_{i\s}^{+}$ the kinetic energy term
has the same form as in \eq{1}, in this representation
the number of double occupied sites is
conserved and the lower bound for the energy of the state with N holes can
be obtained using the representation (\ref{2}) and the theorem (\ref{Haa}).
The energy of isolated $c$ and $d$- particles is $zV$ and their
interaction at nearest neighbor sites is $-V$.
Thus the lower bound for the energy as a function of N is
\be
E_0(N) \geq (2zV - z\max(2t,V) - U)N + UL/2.
\label{bipartite}
\ee
One can see from \eq{bipartite} that
the wave function (\ref{v}) is the ground state at $U<2zV-z\max(2t,V)$.
This ground state is unique apart from a twofold
degeneracy due to the two sublattices and describes the nonmagnetic
insulator. At $V>2t$ we obtain the condition $U<zV$. Since it was shown
that at $U>zV$ the ground state is given by \eq{var}, for a bipartite
lattice at $V>2t$ the ground state is found for arbitrary U.
In the Hubbard model ($X=0$, $V=0$) the function (\ref{var})
is the ground-state wave function only for $U=\infty$. In the generalized
Hubbard model with $X=t$ this ground state is already realized at
$finite$ U with $U>z\max(2t,V)$. Similarly, while in the extended
Hubbard model ($X=0$, $V\neq 0$) the function (\ref{v}) is the ground-state
wave function only for $V=\infty$, in the generalized Hubbard model
with $X=t$ this ground state is already realized at $finite$ V.

\vspace{0.25in}
         {\bf 3.~ Generalized Hubbard model in One dimension.}
\vspace{0.15in}

Let us consider the generalized Hubbard model (\ref{1}) in one dimension.
Before considering the model (\ref{1}) let us study the ground state
multiplicity (the value of the total spin) in the cases when the model is
exactly solvable in one dimension. Namely we consider the model (\ref{1})
at $V=0$ and at $V\neq 0$ in the sector with no double occupied sites.

For  the  one  dimensional  system  the  two different cases can be
considered: 1) the case of an open chain; 2) the case of the
closed chain of finite length. Both cases are equivalent in the
thermodynamic limit ($L\r\infty$). First, consider an open chain or
equivalently the chain of an infinite length. In this case the Nagaoka
theorem is valid at arbitrary number of holes and double occupied
sites since the Hamiltonian (\ref{1}) is  invariant under the
transformation which changes the statistics of holes. In the other
words for an open chain the holes can be considered as a bosons.
That does not mean that the eigenstates with  $S<S_{max}$ cannot be
degenerate in the energy with the state $S=S_{max}$. In fact all the
eigenstates including the ground state are degenerate in the total
spin $S$.  For example let us consider the eigenstates
of the Hamiltonian (\ref{1}) without the double occupied sites.
At $V=0$ in the case when both the numbers of $c$- and
$d$- particles are not equal to zero we get the additional degeneracy of
the eigenstates due to the two species of particles.
Let us seek for the ground state wave function in the following form
\be
\psi(i_1,\ldots i_N|l_1,\ldots l_M) =
\psi_{0}(i_1,\ldots i_N) \phi(\lambda_1,\ldots \lambda_M)
\label{degen}
\ee
where $i_{\a}$ are the coordinates of $c$-  particles
($N=N_c$) and $\lambda_{\beta}$ are the coordinates of the spin bosons
on a "supperlattice" which consists of $L_1=L-N$ lattice sites which are
not occupied by the holes ($L_1=N_e$ is the number of electrons)
\[
\lambda_{\a}=l_{\a}-\sum_{\b=1}^{N}\theta(l_{\a}-i_{\b}),~~~~~~
\a=1,\ldots M.
\]
If $\psi_{0}$ is the eigenstate of the Hamiltonian (\ref{1})
in the sector $S=S_{max}$ then the wave function (\ref{degen}) is the
eigenfunction of the Hamiltonian for an arbitrary function
$\phi(\lambda_1,\ldots\lambda_M )$.
Thus the ground state is degenerate in the total
spin S. Let us turn to the case 2) and see how this degeneracy is resolved
at finite L. Note that unlike the open chain
for the closed chain the spectrum depends on the statistics of particles.
The coordinates on a superlattice can be defined in the
same way by fixing the initial and the final sites of the chain
($l_{\a}=1,\ldots L$ and $\lambda_{\a}=1,\ldots L_1$).
After the substitution (\ref{degen}) we consider
the functions $\psi_0$ and $\phi_0$ extended to the infinite chain and
subjected to the appropriate boundary conditions.
In order to satisfy the periodic boundary conditions for the function
$\psi$ (\ref{degen})  the functions
$\psi_{0}(i_1,\ldots i_N)$ and $\phi(\lambda_1,\ldots \lambda_M)$
should satisfy the following boundary conditions:
\be
\phi(\lambda_1,\ldots\lambda_{\a}+L_1,\ldots \lambda_M)=
\phi(\lambda_1,\ldots\lambda_{\a},\ldots \lambda_M)~~~~
\a=1,\ldots M,
\label{phi}
\ee
and
\be
\psi_{0}(i_1, \ldots i_{\a}+L, \ldots i_N) =
   \exp(iq) \psi_{0}(i_1,\ldots i_{\a}, \ldots i_N),
\label{twist}
\ee
where the boundary conditions for the function $\psi_0$ are determined by
the total momentum $q$ corresponding to the function $\phi$ :
\be
\phi(\lambda_1+1,\ldots \lambda_M+1)=
             e^{iq} \phi(\lambda_1,\ldots \lambda_M).
\label{q}
\ee
The periodic function $\phi$ is the symmetric function of its arguments
which vanishes at $\lambda_{\a}=\lambda_{\b}$. At arbitrary V the function
$\psi_0$ is determined by the set of the momenta $k_{\a}$,
$\a=1,\ldots N$. The energy is
\[
E=-2t\sum_{\a=1}^{N}\cos k_{\a}.
\]
For instance at $V=0$ we have the free fermion determinant
\[
\psi_0=\det_{\a\b} \left[ \exp(ik_{\a}i_{\b}) \right] ,
{}~~~~~~ k_{\a}=2\pi n_{\a}/L,
\]
where $n_{\a}$ are the integers.
The periodic function $\phi$ can be characterized by the set of the momenta
$q_{\a}=2\pi m_{\a}/L_1$ where $m_{\a}$ are integers or half integers
(see below).
An arbitrary number of zeros $q_{\a}=0$ is possible which corresponds to
the value of the difference $S-S^z$. The total momentum
$q=\sum_{\a}2\pi m_{\a}/L_1$.
Thus we obtain the equation for the momenta $k_{\a}$:
\be
k_{\a}=\frac{2\pi}{L}\left(n_{\a}+\sum_{\b=1}^{M}\frac{m_{\b}}{L_1}\right).
\label{moment}
\ee
As a basis in the space of the symmetric functions (\ref{phi})
one can choose the eigenstates of the Hamiltonian of the free hard core
bosons (XX model) on a chain of the length $L_1$.
In this case $m_{\a}$ are integers
(half integers) for $M$- odd (even), and $m_{\a}\neq m_{\b}$ for
$\a\neq\b$. This automatically gives the eigenfunctions
of the Hamiltonian (\ref{1}). However in general this functions are
not the eigenfunctions of the operator of the total spin S. In order to
classify the eigenstates according to their spin  one can choose
the basis given by the eigenstates of the Heisenberg ferromagnet.
In this case an arbitrary number of the momenta $q_{\a}$ can be equal
to zero and the non-zero momenta are determined by the system of the
equations
\be
  e^{iq_{\a}L_1}=(-1)^{M-1}\exp\left( 2i\sum_{\b=1}^{M}\arctg(u_{\a}-u_{\b})
\right),~~~~u_{\a}=\frac{1}{2}\ctg(q_{\a}/2).
\label{Heisen}
\ee
The total spin $S=|L_1/2-M|$, where $M$ is the number of the non-zero
momenta. The total momentum is
$q=\sum_{\a} 2\pi m_{\a}/L_1$ and we obtain the equation
(\ref{moment}) where $n_{\a}$ are integers and $m_{\a}$ are integers
(half integers) for $M$- odd (even) (we assume $L$ to be even).
The same formulas could be obtained starting from the problem of $N_e$
spinless fermions and the hard core bosons (upturned spins) on a
"superlattice" consisting of the lattice sites occupied by the electrons.
The only difference is the number of the momenta $k_{\a}$: $\a=1,...N_e$.
In this representation the same results could be obtained by taking the
limit $U\r\infty$ in the exact Bethe anzatz solution of the 1D Hubbard
model \cite{LW} (for example see ref.\cite{OS}). In fact one can redefine
the quantum numbers according to
$n_{\a}\r n_{\a}-M/2$,~$m_{\a}\r m_{\a}+L_1/2$
to obtain the equation (\ref{moment}) with $n_{\a}$- integers
(half integers) for $M$- even (odd) and $m_{\a}$- integers (half integers)
for $(L_1-M)$- odd (even) in agreement with the results of ref.\cite{OS}.

{}From \eq{moment} the ground state energy as a function
of the total spin can be found. As an example consider the splitting of
the energy levels with $S=S_{max}$ and $S=S_{max}-1$ for an even number
of holes ($L$- is assumed to be even).
Clearly the ground state corresponds to the values $q=0$
and $q=\pi$ respectively and the value $E_0=E_{0}(S_{max}-1)$ is an absolute
ground state energy. Thus we obtain the energy
$E_{0}(S_{max})=E_0\cos(\pi/L)$, where
$E_0=-2t\left(\sin(\pi/L)\right)^{-1}\sin(\pi N/L)$.
For all $S<S_{max}-1$ the minimal energy levels are nearly degenerate
(i.e. the energy splitting is of the higher order in $1/L$ at large
$L$) with the ground state energy $E_0$. The similar picture can be
obtained for $N$ odd. In this case we find $E(S_{max})=E_0$ ($M=0$,
$q=0$) in agreement with the Nagaoka theorem ($N=1$). Clearly the
same procedure (\ref{degen})-(\ref{Heisen})
could be performed at arbitrary $V$ if the double
occupied sites are absent  or at $V=0$ and an arbitrary number of the
double occupied sites. For instance in the first case the Bethe
anzatz equations for the wave function $\psi_0$ (\ref{degen})
with twisted boundary conditions (\ref{twist}) should be used.

Let us consider the model (\ref{1}) at arbitrary $U$ and arbitrary filling
fraction $\bar{n}$. It was proved that in the thermodynamic limit the
eigenstates are degenerate. Thus it is sufficient to consider the
eigenstates with $S=S^z=S_{max}$, which corresponds to the absence of
the spin bosons. First, consider the model at $V=0$.
At $V=0$ the system is equivalent to the system of free fermions with
an extra degeneracy due to the two different species of particles ($c,d$)
with an infinite on-site repulsion. The density of double occupied sites
$\rho=N/L$ is determined by the minimum of the energy of free fermions
with the total density $n_0+2\rho$,
\be
E_0(\rho)/L = -\frac{2t}{\pi}\sin \left[ \pi(n_0+2\rho) \right] +U\rho,
\label{E0}
\ee
where $n_0=1-\bar{n}$ is the concentration of holes in the limit of large U.
Thus at $U>U_c$ where
\[
U_c=4t\cos(\pi n_0),
\]
the transition to the state with no double occupied sites, $\rho=0$,
takes place.
Of course at $n_0\neq 0$ the point $U=U_c$ is not related to the metal -
insulator transition. Away from the half filling the analog of the ground
state (\ref{v}) i.e. the ground state without the single occupied sites is
realized at $U<-4t$. At $V\neq 0$ the model is not exactly solvable in the
sector of Hilbert space with the double occupied sites.
In the previous section it was shown that at arbitrary $V<2t$ and $U>4t$
we have the number of double occupied sites $N=0$
(really $N=0$ at $U>U_c$ where the critical value $U_c<4t$).
At these values the ground state is equivalent to the ground state of the
Heisenberg chain with the anisotropy parameter $V/2t$ in the sector with
the projection of the total spin related to the number of holes.
The spectrum of the anisotropic Heisenberg chain can
be determined exactly with the help the Bethe anzatz. Thus in 1D
the ground  state and the low energy excitations of the model (\ref{1})
can be found exactly at $U>U_c$ and an arbitrary $\bar{n}$.
At $V/2t<1$ and $V=2t$ the spectrum of charge excitation is gapless
and the system is a metal (if the number of holes $n_0>0$). The same is
true for $V/2t>1$ and $n_0<1/2$. Note that away from the half filling
the transition between the state with no double occupied sites and the
state with no single occupied sites analogous to the states (\ref{var})
and (\ref{v}) does not take place at $U=2V$ under the condition $V/2t>1$.
If the concentration of holes is exactly $n_0=1/2$ and $V>2t$ there is a
gap in the spectrum of charge excitations \cite{Gap}. Thus at the
filling fraction $\bar{n}_1=\bar{n}_2=1/4$ the system undergoes
another metal-insulator transition at the point $V=2t$.  In general
the model (\ref{1}) is not exactly solvable in the sector with the
non- zero number of double occupied sites.

Although at the point of metal-insulator transition the ground state
is ferromagnetic for two and three dimensional systems
and degenerate in the total spin in one dimension the existence of
the transition is not connected with the ferromagnetic order. In fact
the metal - insulator transition is general phenomenon in the
models with the kinetic energy term conserving the number of double
occupied sites (\ref{1}).  For example one can study the metal -
insulator transition in the one dimensional models with an
antiferromagnetic coupling which
are exactly solvable in the absence of double occupied sites
\cite{Slott},\cite{S},\cite{B}. Apart from the term $\sim X$
($X=t$) and the on-site repulsion $\sim U$ these models include the
interaction of the form
$J\sum (S_i S_j-{1\over 4}n_i n_j)$ or
$J\sum (S_i S_j+{3\over 4}n_i n_j)$ at $J=2t$.
Although these models are not exactly solvable at $N\neq 0$ at the
half filling the existence of metal-insulator transition can be
shown and the critical value of U can be found exactly
$U_c=2zt\ln 2$.
The model which is exactly solvable at arbitrary N was proposed in
ref.\cite{Korepin}. The Hamiltonian is the sum of the Hamiltonian of
the tJ- model (modified to include the double occupied sites) and the
permutation term of the form
\[
2t\sum_{<ij>}(c_{i}^{+}c_{j}d_{j}^{+}d_i+c_{j}^{+}c_{i}d_{i}^{+}d_j).
\]
One can also change the sign $J\r -J$ to obtain the integrable model
with the ferromagnetic ground state. The ground state of this model
is equivalent to the ground state of the tJ- model \cite{Slott} with
the up- and down- spin electrons replaced by $c$- and $d$- particles
(that corresponds to the $SU(2)$ "$\eta$- spin" symmetry of
ref.\cite{Yang}).
The concentration of $d$- particles should be found from the
condition of minimum of the energy.

\vspace{0.25in}
                 {\bf 4.~ Conclusion.}
\vspace{0.15in}

In conclusion, for the model (\ref{1})
it was shown that at the half filling and $U>z\max(2t,V)$
the ground state is given by \eq{var}.
The problem with one hole and one double occupied site in the
ferromagnetic background was solved. We proved that it is the lowest
energy state in the sector of Hilbert space with one empty and one
double occupied sites.
We established that at $V<2t$ the critical value of U corresponding to
the point of metal-insulator transition $U_c=2zt$.
Under the assumption about the stability of the ferromagnetic state
at finite concentration of holes the density of holes was found at
$V<2t$ and $U\r U_c$.
Finally, for a $bipartite$ lattice with constant number of nearest neighbors
at $U<2zV-z\max(2t,V)$ the exact ground state wave function is given by
\eq{v}. At $V>2t$ the ground state is found exactly at arbitrary value of
the parameter $U$. The transition between the states (\ref{var}) and
(\ref{v}) occurs at $U=zV$. At $V<2t$ the exact ground state is found at
$U>2zt$ (\ref{var}) and $U<2zV-2zt$ (\ref{v}).
For the one dimensional model at $V=0$ the exact solution was
presented. We have also studied the dependence of the energy on the
total spin.
Recently the part of the results obtained in Section 3 was independently
obtained in ref.\cite{deBoer}.

This work was supported, in part, by the Weingart foundation through
a cooperative agreement with the Department of Physics at UCLA.

\vspace{0.5in}
\begin{center}
                     {\bf Appendix}
\end{center}

Following Nagaoka \cite{N}, we introduce a set of orthogonal and normalized
many-body wave functions which completely span the Hilbert space.
We use the representation (\ref{2}) in terms of the operators of holes
$c_{i}^{+}=c_{2i}$, double occupied sites $d_{i}^{+}=c_{1i}^{+}$ and
the overturned spins (hard core bosons) $b_{i}^{+}=c_{1i}^{+}c_{2i}$
starting from the ferromagnetic state $\sum_{i}c_{2i}^{+}|0>$.
Because of the Fermi statistics we have to be cautious about the order of
$c$- and $d$- operators in the definition of the states $\a,\b$.
Let for each site $i$ of the lattice $Ri$ to be an integer number
$Ri=1,\ldots L$. We can define the following order among the lattice sites.
Setting up the coordinate system we assign a pair of integer coordinates
$(i_{x},i_y)$ to each site of the lattice. If for the two sites $i$, $j$
$i_x<j_x$ then we define $Ri<Rj$. When $i_x=j_x$ the order is determined by
their $y$ coordinates: $Ri<Rj$ for $i_y<j_y$. With this order we now
introduce the states
\[
|\a>= \ck_{i_1}\ldots \ck_{i_N} \, \dk_{j_1}\ldots \dk_{j_N} \,
\bk_{l_1}\ldots \bk_{l_M}|0>,
\]
where the order of the operators for a given set of the lattice sites
$\a=(i_1...i_N|j_1...j_N|l_1...l_M)$ is given by the condition
\[
Ri_1 < \ldots < Ri_N,~~ Rj_1 < \ldots < Rj_N,~~ Rl_1 < \ldots < Rl_M.
\]
With this definition the non-diagonal matrix elements
$H_{\a\b}=<\a|\hat{H}|\b>$ of the
Hamiltonian (\ref{2}) are equal to $-t$ when one hole (double occupied site)
changes order with an even number of holes (double occupied sites).
In the opposite case the matrix elements $H_{\a\b}$ are equal to $+t$.
Clearly for the Bose statistics the matrix elements would be equal to $-t$.
Since $c$- and $d$- particles are the two distinct species of
fermions one can use the prescription given by the equation for $|\a>$.
For $N=1$ the non-diagonal matrix elements are also equal to $-t$
which justifies our statement that the Fermi statistics of $c$- and $d$-
particles is not important for the case of one hole and one double occupied
site. Thus for the model (\ref{1}) the Nagaoka proof \cite{N} can be used
without modifications for one hole and one double occupied site.


\begin{thebibliography}{99}

\bibitem{H}
J.Hubbard, Proc.R.Soc.London A 276 (1963) 238. \\
M.C.Gutzwiller, Phys.Rev.Lett. 10 (1963) 159.

\bibitem{HM}
F.Marsiglio, J.E.Hirch, Phys.Rev.B 41 (1990) 6435. \\
J.E.Hirch, F.Marsiglio, Phys.Rev.B 41 (1990) 2049; Physica C
162-164 (1989) 591.

\bibitem{VS}
R.Strack, D.Vollhardt, Phys.Rev.Lett. 70 (1993) 2637.

\bibitem{Others}
D.Baeriswyl, P.Horsch, K.Maki, Phys.Rev.Lett. 60 (1988) 70.\\
D.K.Campbell, J.T.Gammel, E.Y.Loh,  Phys.Rev.B 42 (1990) 475.\\
P.Vries, K.Michielsen, H.Raedt, Phys.Rev.Lett. 70 (1993) 2463.\\
J.Appel, M.Grodzicki, F.Paulsen, Phys.Rev.B 47 (1993) 2812.

\bibitem{Korepin}
F.H.L.Essler, V.E.Korepin, K.Schoutens, Phys.Rev.Lett. 68 (1992) 2960;
70 (1993) 73.

\bibitem{MR}
A.Montorsi, M.Rasetti, Phys.Rev.Lett. 66 (1991) 1383.

\bibitem{Mott}
N.F.Mott, {\it Metal-Insulator Transitions}
(Taylor \& Frances, London, 1974).

\bibitem{BR}
W.F.Brinkman, T.M.Rice, Phys.Rev.B 2 (1970) 4302. \\
G.Kotliar, A.E.Ruckenstein, Phys.Rev.Lett. 57 (1986) 1362.

\bibitem{N}
Y.Nagaoka, Phys.Rev. 147 (1966) 127.

\bibitem{A}
B.S.Shastry, H.R.Krishnamurthy, P.W.Anderson, Phys.Rev.B 41 (1990) 2375.\\
A.Barbieri, J.A.Riera, A.P.Young, Phys.Rev.B 41 (1990) 11697. \\
T.Kopp, A.E.Ruckenstein, S.Schmitt-Rink, Phys.Rev.B 42 (1990) 6850. \\
R.Richmond, G.Rickayzen, J.Phys.C 2 (1969) 528. \\
A.G.Basile, V.Elser, Phys.Rev.B 41 (1990) 4842.

\bibitem{Ovch}
A.A.Ovchinnikov, Mod.Phys.Lett.B 7 (1993) 1397.

\bibitem{Bloom}
P.Bloom, Phys.Rev.B 12 (1975) 125.

\bibitem{Cyrot}
M.Cyrot, Physica 91B (1977) 141.

\bibitem{Raedt}
P.de Vries, K.Michielsen, H.De Raedt, Phys.Rev.Lett. 70 (1993) 2463.

\bibitem{LW}
E.H.Lieb, F.Y.Wu, Phys.Rev.Lett. 20 (1968) 1445.

\bibitem{OS}
M.Ogata, H.Shiba, Phys.Rev.B 41 (1990) 2326.

\bibitem{Gap}
J.Cloizeaux, M.Gaudin, J.Math.Phys. 7 (1966) 1384.

\bibitem{Slott}
P.Schlottmann, Phys.Rev.B 36 (1987) 5177.

\bibitem{S}
B.Sutherland, Phys.Rev.B 12 (1975) 3795.

\bibitem{B}
P.A.Bares, G.Blatter, M.Ogata, Phys.Rev.B 44 (1991) 130.

\bibitem{Yang}
C.N.Yang, Phys.Rev.Lett. 63 (1989) 2144. \\
C.N.Yang, S.Zhang, Mod.Phys.Lett.B 4 (1990) 759.

\bibitem{deBoer}
Jan de Boer, V.E.Korepin, A.Schadschneider, Preprint ITP-SB-94-18.

\end{thebibliography}
\end{document}